\documentclass[prb,floatfix,twocolumn,showpacs,amsmath,amssymb]{revtex4}
\usepackage{graphicx}% Include figure files
\usepackage{dcolumn}% Align table columns on decimal point
\usepackage{bm}% bold math
\begin{document}

\title{Role of backflow correlations for the non-magnetic phase of the 
$t{-}t^\prime$ Hubbard model}
\author{Luca F. Tocchio,$^{1,2}$ Federico Becca,$^{1,2}$ 
Alberto Parola,$^{3}$ and Sandro Sorella,$^{1,2}$}
\affiliation{
$^{1}$ International School for Advanced Studies (SISSA), 
Via Beirut 2, I-34014 Trieste, Italy \\
$^{2}$ CNR-INFM-Democritos National Simulation Centre, Trieste, Italy. \\
$^{3}$ Dipartimento di Fisica e Matematica, Universit\`a dell'Insubria,
Via Valleggio 11, I-22100 Como, Italy
}

\date{\today} 

\begin{abstract}
We introduce an efficient way to improve the accuracy of projected wave 
functions, widely used to study the two-dimensional Hubbard model. 
Taking the clue from the backflow contribution, whose relevance has been 
emphasized for various interacting systems on the continuum, we consider 
many-body correlations to construct a suitable approximation for the ground 
state at intermediate and strong couplings. In particular, we study the phase 
diagram of the frustrated $t{-}t^\prime$ Hubbard model on the square lattice 
and show that, thanks to backflow correlations, an insulating and 
non-magnetic phase can be stabilized at strong coupling and sufficiently 
large frustrating ratio $t^\prime/t$. 
\end{abstract}


\maketitle

{\it Introduction}.
Recently, the interest in the role of frustrating interactions in electronic 
systems has considerably increased since in this regime new exotic phases may 
appear. Many experiments suggest the possibility to have disordered phases 
down to very low-temperatures (much smaller than what one would expect from a 
mean-field approach) or even to zero temperature. Such phases are generically 
called spin liquids. In this respect, the organic molecular materials 
$\kappa$-(ET)$_2$X, X being a monovalent anion,~\cite{kanoda1,kanoda2} 
represent an interesting example, since they show a particularly rich phase 
diagram. In the conducting layers, ET molecules are strongly dimerized and 
form a two-dimensional (2D) triangular lattice. Since the valence of each ET 
dimer is $+1$, the conduction band is half filled. By acting with an external 
pressure, it is possible to vary the ratio between the on-site Coulomb 
repulsion and the bandwidth, driving the system through a metal-insulator 
transition (MIT). 

The minimal model to describe the physics of correlated electrons is the 
Hubbard model 
\begin{equation}\label{hubbard}
{\cal H}=-\sum_{i,j,\sigma} t_{ij} c^{\dagger}_{i,\sigma} c_{j,\sigma} + H.c.
+U \sum_{i} n_{i,\uparrow} n_{i,\downarrow},
\end{equation}
where $c^{\dagger}_{i,\sigma} (c_{i,\sigma})$ creates (destroys) an electron 
with spin $\sigma$ on site $i$, 
$n_{i,\sigma}=c^{\dagger}_{i,\sigma}c_{i,\sigma}$, $t_{ij}$ is the hopping
amplitude, that determines the bandwidth, and $U$ is the on-site Coulomb
repulsion. In this work, we focus our attention on the half-filled case with
$N$ electrons on $N$ sites and consider the square lattice with both nearest-
and next-nearest-neighbor hoppings, denoted by $t$ and $t^\prime$, 
respectively. This model represents the prototype for frustrated electronic 
materials,~\cite{hirsch} and, recently, it has been widely studied by 
different numerical techniques, with contradictory 
outcomes.~\cite{imada1,imada2,imada3,ogata,tremblay} Here we present the 
results for the zero-temperature phase diagram, obtained by using projected 
wave functions, which only after considering backflow correlations are
accurate enough to describe the highly-correlated regime.

{\it The variational approach}.
Variational wave functions for the unfrustrated Hubbard model,
describing the antiferromagnetic phase, can be constructed by considering the 
ground state $|AF\rangle$ of a mean-field Hamiltonian containing a band 
contribution and a magnetic term 
${\cal H}_{AF}=\Delta_{AF} \sum_j e^{i{\bf Q} \cdot {\bf R}_j} S_j^x$, where
$S_j^x$ is the $x$ component of the spin operator
${\bf S}_j=(S_j^x,S_j^y,S_j^z)$. In order to have the correct spin-spin 
correlations at large distance, we have to apply a suitable long-range spin 
Jastrow factor, namely $|\Psi_{AF}\rangle = {\cal J}_s |AF\rangle$, with
${\cal J}_s=\exp [-\frac{1}{2} \sum_{i,j} u_{i,j} S_i^z S_j^z ]$, 
which governs spin fluctuations orthogonal to the magnetic field
$\Delta_{AF}$.~\cite{becca}

On the other hand, spin-liquid (i.e., disordered) states can be constructed 
by considering the ground state $|BCS\rangle$ of a BCS Hamiltonian and then 
applying to it the so-called Gutzwiller projector, 
$|RVB\rangle = {\cal P}_G |BCS\rangle$, where 
${\cal P}_G = \prod_i (1-g n_{i,\uparrow} n_{i,\downarrow})$ and 
$g=1$.~\cite{anderson,gros}
In pure spin models, where the $U$ is {\it infinite} and charge fluctuations 
are completely frozen, these kind of states can be remarkably accurate and 
provide important predictions on the stabilization of 
disordered spin-liquid ground states.~\cite{capriotti,yunoki} 
However, whenever $U/t$ is finite, the variational state must also contain 
charge fluctuations. In this regard, the simplest generalization of the
Gutzwiller projector with $g<1$, that allows doubly occupied sites, is known 
to lead to a metallic phase.~\cite{shiba} In order to obtain a Mott insulator 
with no magnetic order, it is necessary to consider a sufficiently long-range
Jastrow factor ${\cal J}=\exp [-\frac{1}{2} \sum_{i,j} v_{i,j} n_i n_j ]$,
$n_i=\sum_\sigma n_{i,\sigma}$ being the local density.~\cite{capello}
Nevertheless, the accuracy of the resulting wave function 
$|\Psi_{BCS}\rangle = {\cal J} |BCS\rangle$ can be rather poor in 2D for large 
on-site interactions,~\cite{capellophd} especially in the presence of 
frustration (see below). Therefore, other contributions beyond the Jastrow 
factor must be included. 

An alternative way, suitable to describe the strong-coupling regime, is to 
start from the fully-projected state $|RVB\rangle$ and then apply the unitary 
transformation, that was introduced long time ago to connect the Heisenberg 
and the Hubbard models,~\cite{girvin} namely
$|\Psi_S\rangle=\exp(iS)|RVB\rangle$. This kind of approach is rather 
difficult to implement for large clusters, since, in contrast to the Jastrow 
term, $S$ is non-diagonal in the natural basis $|x\rangle$ where the electrons 
with spins quantized along $z$ occupy the lattice sites. The expression valid 
for strong coupling, i.e., $(1+iS)|RVB\rangle$, is clearly not accurate for 
large system sizes, since it can allow a single doubly-occupied site at most.
In this respect, accurate results for small clusters can be also obtained by
performing  one Lanczos step, 
$(1+\alpha {\cal H})|\Psi_{BCS}\rangle$,~\cite{becca} or by considering
$\exp(hK) {\cal P}_G|BCS\rangle$ (where $h$ and $g$ are variational parameters 
and $K$ is the hopping Hamiltonian).~\cite{baeriswyl}

%%%%%%%%%%%%%%%%%%%%%%%%%%%%%%%%%%%%%%%%%%%%%%%%%%%%%%%%%%%%%%%%%%%%%%%
\begin{figure}
\includegraphics[width=\columnwidth]{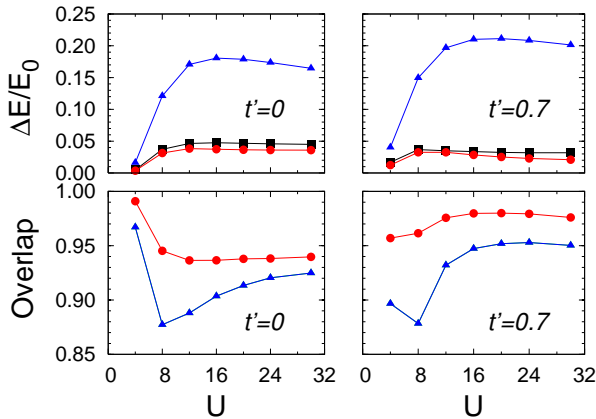}
\caption{\label{fig:18site}
(Color online) Results for 18 electrons on 18 sites as a function of $U/t$. 
Upper panels: Accuracy of energy $\Delta E=(E_0-E_v)$, $E_v$ and $E_0$ being 
the variational and the exact values, respectively. Lower panels: Overlap
between the exact ground state and the variational wave functions. The
BCS state with long-range Jastrow factor is denoted by blue triangles, 
the BCS state with backflow correlations and Jastrow term  by red circles. 
The results for $\Delta E$ considering one Lanczos step upon the BCS state, 
i.e., $(1+\alpha {\cal H})|\Psi\rangle$, are also shown (black squares).}
\end{figure}
%%%%%%%%%%%%%%%%%%%%%%%%%%%%%%%%%%%%%%%%%%%%%%%%%%%%%%%%%%%%%%%%%%%%%%

{\it The backflow wave function}. 
In order to improve in a size-consistent way the previous wave functions
$|\Psi_{AF}\rangle$ and $|\Psi_{BCS}\rangle$, we want to modify the 
single-particle orbitals,~\cite{noteph} in the same spirit of the backflow
correlations, which have been proposed long time ago by Feynman and 
Cohen, to obtain a quantitative description of the roton excitation in liquid
Helium.~\cite{feynman} The backflow has been implemented within quantum Monte 
Carlo calculations to study bulk liquid $^3$He,~\cite{schmidt1,schmidt2} and 
used to improve the description of the electron jellium both in two and three 
dimensions.~\cite{ceperley1,ceperley2} More recently, it has been applied 
to metallic Hydrogen.~\cite{ceperley3} Originally, the backflow term 
corresponds to consider fictitious coordinates of the electrons 
${\bf r}^b_{\alpha}$, which depend upon the positions of the other particles, 
so to create a return flow of current: 
\begin{equation}
{\bf r}^b_{\alpha} = {\bf r}_{\alpha} + \sum_{\beta} 
\eta_{\alpha,\beta} [x]  \left ( {\bf r}_{\beta} - {\bf r}_{\alpha} \right ),
\end{equation}
where ${\bf r}_{\alpha}$ are the actual electonic positions and 
$\eta_{\alpha,\beta}[x]$ are variational parameters depending in principle 
on all the electronic  coordinates, namely on the many-body configuration 
$|x\rangle$. The variational wave function is then constructed by means of 
the orbitals calculated in the new positions, i.e., $\phi({\bf r}^b_{\alpha})$. 
Alternatively, the backflow can be introduced by considering a linear expansion
of each single-particle orbital: 
\begin{equation}\label{key}
\phi_k({\bf r}^b_{\alpha}) \simeq \phi_k^b({\bf r}_{\alpha}) \equiv 
\phi_k({\bf r}_{\alpha}) + \sum_{\beta} c_{\alpha,\beta} [x] \; 
\phi_k({\bf r}_{\beta}),
\end{equation}
where $c_{\alpha,\beta}[x]$ are  suitable coefficients.  
The definition~(\ref{key}) is particularly useful in lattice models, where the
coordinates of the particles may assume only discrete values. 
In particular, in the Hubbard model, the form of the new ``orbitals'' can be 
fixed by considering the $U \gg t$ limit, so to favor a recombination of 
neighboring charge fluctuations (i.e., empty and doubly-occupied sites):
\begin{equation}\label{backlattice1}
\phi_k^b({\bf r}_{i,\sigma}) \equiv \epsilon \phi_k({\bf r}_{i,\sigma})
+ \eta \sum_j t_{ij} \left( D_i H_j \right) \phi_k({\bf r}_{j,\sigma}),
\end{equation}
where we used the notation that 
$\phi_k({\bf r}_{i,\sigma})= \langle 0|c_{i,\sigma}|\phi_k\rangle$, being 
$|\phi_k\rangle$ the eigenstates of the mean-field Hamiltonian,
$D_i=n_{i,\uparrow}n_{i,\downarrow}$, $H_i=h_{i,\uparrow}h_{i,\downarrow}$, 
with $h_{i,\sigma}=(1-n_{i,\sigma})$,
so that $D_i$ and $H_i$ are non zero only if the site $i$ is doubly occupied
or empty, respectively; finally $\epsilon$ and $\eta$ are variational 
parameters (we can assume that $\epsilon=1$ if $D_iH_j=0$).
As a consequence, already the determinant part of the wave function includes 
correlation effects, due to the presence of the many body operator $D_i H_j$. 
The previous definition of the backflow term preserves the spin SU(2) 
symmetry. A further generalization of the new ``orbitals'' can be made, 
by taking all the possible virtual hoppings of the electrons:
\begin{eqnarray}\label{backlattice2}
&&\phi_k^b({\bf r}_{i,\sigma}) \equiv \epsilon \phi_k({\bf r}_{i,\sigma})
+ \eta_1 \sum_j t_{ij} \left( D_i H_j \right) \phi_k({\bf r}_{j,\sigma})
+ \nonumber \\
&& \eta_2 \sum_j t_{ij} \left( n_{i,\sigma} h_{i-\sigma}
n_{j,-\sigma} h_{j\sigma} \right) \phi_k({\bf r}_{j,\sigma}) + \nonumber \\
&& \eta_3 \sum_j t_{ij} \left( D_i n_{j,-\sigma} h_{j\sigma}
+n_{i,\sigma} h_{i-\sigma} H_j \right) \phi_k({\bf r}_{j,\sigma}),
\end{eqnarray}
where $\epsilon$, $\eta_1$, $\eta_2$, and $\eta_3$ are variational parameters.
The latter two variational parameters are particularly important for the 
metallic phase at small $U/t$, whereas they give only a slight improvement
of the variational wave function in the insulator at strong coupling.
The definition Eq.~(\ref{backlattice2}) may break the SU(2) symmetry,
however, the optimized wave function always has a very small value of the
total spin square, i.e., $\langle S^2 \rangle \sim 0.001$ for 50 sites.
All the parameters of the wave function (contained in the mean-field 
Hamiltonian, in the Jastrow term, and in the backflow term) can be 
optimized by using the method of Ref.~\onlinecite{yunoki}. Finally, the 
variational results can be compared with more accurate (and still variational) 
ones obtained by Green's function Monte Carlo,~\cite{nandini} 
implemented with the so-called fixed-node (FN) approximation.~\cite{ceperleyfn}
 
{\it Results}.
Let us start by considering the comparison of the variational results with the
exact ones on the 18-site cluster at half filling. In Fig.~\ref{fig:18site}, 
we show the accuracy of the variational BCS state (with and without backflow
correlations) and the overlap with the exact ground state for two values of 
the frustrating ratio, i.e.,  $t^\prime/t=0$ and $0.7$. The backflow term is 
able to highly improve the accuracy both for weak and strong couplings.
We also notice that backflow correlations are more efficient than
applying one Lanczos step, i.e., $(1+\alpha {\cal H})|\Psi_{BCS}\rangle$, that 
was used in previous calculations.~\cite{becca} The overlap between the exact 
ground state and the backflow state remains very high, even for large $U$,
and the improvement with respect to the BCS state is crucial, especially in 
the frustrated regime. 

%%%%%%%%%%%%%%%%%%%%%%%%%%%%%%%%%%%%%%%%%%%%%%%%%%%%%%%%%%%%%%%%%%%%%%%
\begin{figure}
\includegraphics[width=\columnwidth]{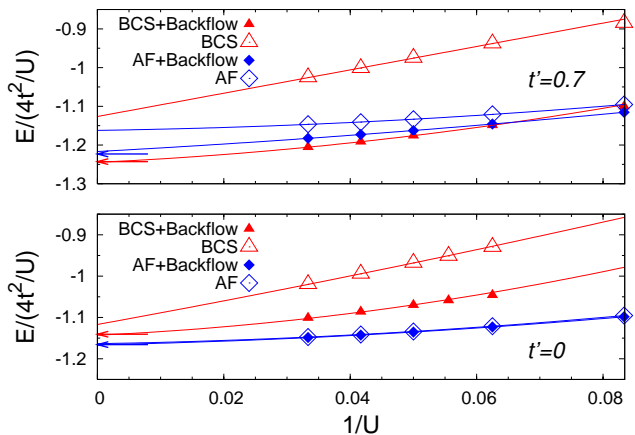}
\caption{\label{fig:extrapol}
(Color online) Variational energies per site (in unit of $J=4t^2/U$) for the 
BCS state with long-range Jastrow factor, with and without backflow 
correlations, and 98 sites. The results for the wave function with 
antiferromagnetic order and no BCS pairing are also shown. Arrows indicate 
the variational results obtained by applying the full Gutzwiller projection 
to the mean-field states for the corresponding Heisenberg models.}
\end{figure}

\begin{figure}
\includegraphics[width=\columnwidth]{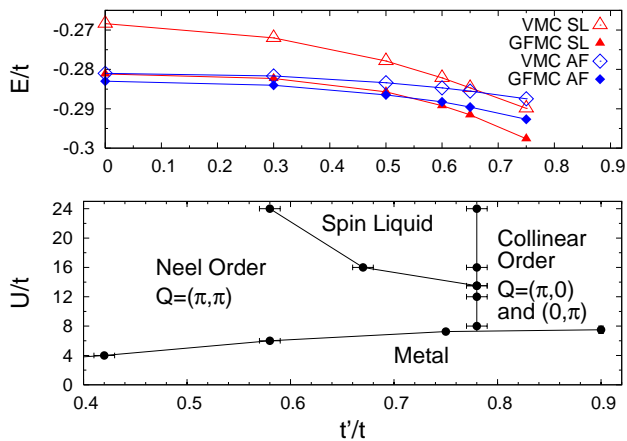}
\caption{\label{fig:phase}
(Color online) Lower panel: Phase diagram of the frustrated $t{-}t^\prime$ 
Hubbard model, as obtained by comparing the variational energies of the 
backflow wave functions. Upper panel: Comparison between the variational 
energies per site and the FN ones for $U/t=16$ and 98 sites.}
\end{figure}

\begin{figure}
\includegraphics[width=\columnwidth]{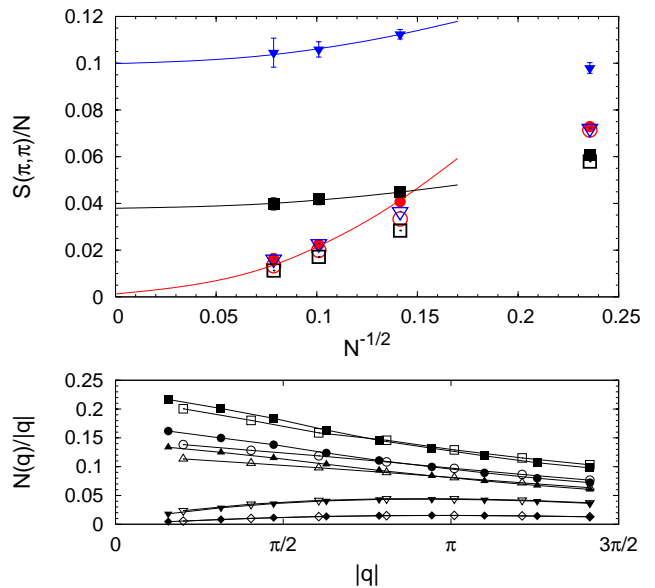}
\caption{\label{fig:nq}
(Color online) Upper panel: Variational (empty symbols) and FN (full symbols)
results for $S(\pi,\pi)$ divided by $N$. Both calculations have been done by 
using the projected BCS wave function; $U/t=16$ and $t^\prime/t=0$ (triangles),
$U/t=24$ and $t^\prime/t=0.7$ (circles), and $U/t=8$ and $t^\prime/t=0.75$ 
(squares). Lines are guides to the eye. 
Lower panel: Variational results for $N(q)$ divided by $|q|$ for 
98 (empty symbols) and 162 (full symbols) sites and $t^\prime/t=0.75$; 
from top to bottom: $U/t=4$, $6$, $7$, $8$, and $16$.}
\end{figure}
%%%%%%%%%%%%%%%%%%%%%%%%%%%%%%%%%%%%%%%%%%%%%%%%%%%%%%%%%%%%%%%%%%%%%%%

Backflow correlations remain efficient also for larger sizes and provide much
lower energy than the Lanczos step wave function, e.g., for $98$ sites with
$U/t=20$ and $t^\prime/t=0.7$, the energy per site with the backflow
wave function is $E_b/t=-0.2352(1)$, while the one with one Lanczos step is
$E_{ls}/t=-0.2310(1)$ (for $18$ sites they are $E_b/t=-0.23741$ and
$E_{ls}/t=-0.23566$). The FN energy obtained with the backflow state is
$E_{FN}/t=-0.2395(1)$, rather close to our estimation of the exact value 
(based upon an extrapolation obtained with zero and one Lanczos step) that
is $E/t \sim -0.246$.  

By increasing $U/t$, the variational energy extrapolates to the one obtained 
by taking the fully-projected state $|RVB\rangle$ in the spin model. 
On the contrary, without using backflow terms, the energy of the BCS state, 
even in presence of a fully optimized Jastrow factor, is few hundredths of 
$J=4t^2/U$ higher than the expected value, see Fig.~\ref{fig:extrapol}. 
Moreover, whenever frustration is large enough, backflow correlations are 
useful also in the antiferromagnetic state $|\Psi_{AF}\rangle$, while for 
$t^\prime=0$ they are not necessary to extrapolate correctly to the value of 
the spin model, see Fig.~\ref{fig:extrapol}.

In order to draw the ground-state phase diagram of the $t{-}t^\prime$ Hubbard
model, we consider three different wave functions with backflow correlations: 
Two antiferromagnetic states $|\Psi_{AF}\rangle$ with ${\bf Q}=(\pi,\pi)$ 
and ${\bf Q}=(\pi,0)$, relevant for small and large $t^\prime/t$, and the 
non-magnetic state $|\Psi_{BCS}\rangle$. The variational phase diagram is 
reported in Fig.~\ref{fig:phase}. The first important outcome is that,
without backflow terms, the energies of the spin-liquid wave function are 
{\it always} higher than those of the magnetically ordered states, for any 
value of frustration $t^\prime/t$. Instead, by inserting backflow correlations,
a spin-liquid phase can be stabilized at large enough $U/t$ and 
frustration (see also Fig.~\ref{fig:extrapol}). 
The small energy difference between the pure variational and the FN energies 
demonstrates the accuracy of the backflow states, see Fig.~\ref{fig:phase}.
Notice that $|\Psi_{AF}\rangle$ and $|\Psi_{BCS}\rangle$ have different nodal 
surfaces, implying different FN energies.

For small Coulomb repulsion and finite $t^\prime/t$ the static density-density 
correlations $N(q)=\langle n_{-q}n_{q} \rangle$ (where $n_q$ is the Fourier 
transform of the local density $n_i$) have a linear behavior for $|q| \to 0$,
typical of a conducting phase. A very small superconducting parameter with 
$d_{x^2-y^2}$ symmetry can be stabilized, suggesting that long-range pairing 
correlations, if any, are tiny. By increasing $U/t$, a MIT is found and $N(q)$ 
acquires a quadratic behavior in the insulating phase, indicating a vanishing 
compressibility. This behavior does not change when considering the FN 
approach, though the metal-insulator transition may be slightly shifted.
In Fig.~\ref{fig:nq}, we show the variational results for $N(q)$ as a function 
of $U/t$ for $t^\prime/t=0.75$. The insulator just above the transition is 
magnetically ordered and the variational wave function has a large 
$\Delta_{AF}$; the transition is likely to be first order. By further 
increasing $U/t$, there is a second transition to a disordered insulator. 
Indeed, for $U/t \gtrsim 14$, the energy of the BCS wave function 
becomes lower than the one of the antiferromagnetic state. In this respect, 
the key ingredient to have such an insulating behavior is the presence of a 
singular Jastrow term $v_q \sim 1/q^2$, that turns a BCS superconductor into a 
Mott insulator.~\cite{capello} In contrast to previous 
investigations,~\cite{imada1,imada2,imada3,ogata,tremblay} for intermediate
on-site couplings, our calculations indicate the possibility to have a direct 
(first-order) transition between two magnetic states, see Fig.~\ref{fig:phase}.

In order to verify the magnetic properties obtained within the
variational approach, we can consider the static spin-spin correlations
$S(q)=\langle S^z_{-q}S^z_{q} \rangle$ over the FN ground state.
Although the FN approach may break the SU(2) spin symmetry, favoring a spin 
alignment along the $z$ axis (this is what we find for small lattices, by a 
direct comparison with exact results), $S(q)$ is particularly simple to 
evaluate within this approach,~\cite{nandini} and it gives important insights 
into the magnetic properties of the ground state.
In Fig.~\ref{fig:nq}, we report the comparison between the variational and the
FN results by considering the non-magnetic state $|\Psi_{BCS}\rangle$. 
Remarkably, in the unfrustrated case, where antiferromagnetic order is 
expected, the FN approach is able to increase spin-spin correlations at 
$q=(\pi,\pi)$, even by considering the non-magnetic wave function to fix the
nodes. A finite value of the magnetization is also plausible in the insulating 
region just above the metallic phase at strong frustration 
(i.e., $t^\prime/t \sim 0.75$), confirming the pure variational calculations. 
On the contrary, by increasing the electron correlation, the FN results change 
only slightly the variational value of $S(\pi,\pi)$, indicating the stability 
of the disordered state.

In conclusion, we have introduced a novel wave function, that highly improves 
the accuracy of the projected states, used so far. Our variational ansatz is
particularly useful to describe non-magnetic phases, that can be stabilized in
the strong-coupling regime of the $t{-}t^\prime$ Hubbard model on the square
lattice.

We acknowledge partial support from CNR-INFM.


\begin{thebibliography}{99}
\bibitem{noteph} After the particle-hole transformation on down spins
   $c^\dagger_{i,\downarrow} \to c_{i,\downarrow}$, the BCS Hamiltonian 
   conserves the particle number and it is possible to define single-particle 
   states (with no definite spin along $z$).
\end{thebibliography}
\end{document}